# Current practice in software development for computational neuroscience and how to improve it

Marc-Oliver Gewaltig[1] and Robert Cannon[2]

1. Blue Brain Project, Ecole Polytechnique Federal de Lausanne, 1015 Lausanne, Switzerland
2. Textensor Ltd., Edinburgh, EH2 2JR, United Kingdom

## Abstract

Almost all research work in computational neuroscience involves software. As researchers try to understand ever more complex systems, there is a continual need for software with new capabilities. Because of the wide range of questions being investigated, new software is often developed rapidly by individuals or small groups. In these cases, it can be hard to demonstrate that the software gives the right results. Software developers are often open about the code they produce and willing to share it, but there is little appreciation among potential users of the great diversity of software development practices and end results, and how this affects the suitability of software tools for use in research projects. To help clarify these issues, we have reviewed a range of software tools and asked how the culture and practice of software development affects their validity and trustworthiness.

We identified four key questions that can be used to categorize software projects and correlate them with the type of product that results. The first question addresses what is being produced. The other three concern why, how, and by whom the work is done. The answers to these questions show strong correlations with the nature of the software being produced, and its suitability for particular purposes. Based on our findings, we suggest ways in which current software development practice in computational neuroscience can be improved and propose checklists to help developers, reviewers and scientists to assess the quality whether particular pieces of software are ready for use in research.

## Introduction

Like most areas of scientific investigation, neuroscience is increasingly dependent on software. Software is used for recording and analyzing experimental data. It is also used in computational models that make it possible to perform detailed quantitative studies of phenomena that are too intricate or complex to be elucidated by abstract reasoning or mathematics alone. In some cases, existing tools are perfectly adequate. In others, the only way to provide required functionality is to write the software from scratch. Many studies involve a mix of the two approaches: existing tools are combined with custom software implementing new models, or combining old tools in new ways. This leads to continual production of software.

Despite great diversity in the nature of the software created, and in the reasons for it being written, publications involving the use of simulation software tend to treat it all in the same way. This can lead to misunderstanding and disappointment when it turns out that software used for a particular study is not sufficiently well written, or well documented to be used or extended by others.

The phenomenon is not unique to neuroscience. In proposing guidelines for scientific software development, Baxter et al. [1] wrote of their

> *"...collective, heartbreaking experiences watching wheels reinvented, finding dead or unusable programs, and, worse, inheriting rancid and labyrinthine code bases."*

Although this is clearly disappointing, it may also be inevitable. Of course, it is possible to develop software in a highly structured and disciplined way, in which all the output is of a very high standard. However, this typically requires large teams with rather low output per developer. In science, developers often work alone,, and are learning their skills as they go along. In these conditions, it is natural that the results are voluminous but of very variable quality.

The greatest challenge with this kind of disparate, and ad-hoc development model, is to ensure that the software used in research studies is actually doing what the developers intended. When Donoho et al. [2] examined the methods used to validate scientific software they found that:

> *"The vast body of results being generated by current computational science practice suffer a large and growing credibility gap: it is impossible to verify most of the computational results shown in conferences and papers."*

They went on to conclude that:

> *"[C]urrent computational science practice does not generate routinely verifiable knowledge."*

These issues came to media prominence in 2009, with the exposure of source code from the University of East Anglia's Climate Research Unit, used for processing global temperature data. The code contained frank, and frequently critical comments, from a software developer. When questioned by the press, professional software engineers expressed views ranging from resigned acknowledgment to outright incredulity at the state of the code [3]. In the light of findings by Donoho et al. [2] and Baxter et al. [1], this situation will be unsurprising to many scientific software developers. They are likely to be relieved, however, that their own code is not subject to such scrutiny, and that it is not being used to generate data that informs global decisions.

There are numerous documented cases where scientific software has fallen short, leading to erroneous conclusions with significant consequences. For example, Post and Votta [4] report that the U.S. withdrew from the International Thermonuclear Experimental Reactor (ITER) project in 1998, on the basis of preliminary and, as was later found, incorrect simulations. The U.S is currently trying to rejoin ITER. More recently, Miller [5] reported in *Science* that five high-profile articles (three in *Science* and two in other journals) had to be retracted because of an error in analysis software that the authors had "inherited" from another laboratory. It seems likely that these documented cases only scratch the surface of a much bigger problem, and that the majority of errors due to faulty and unreliable software remain undetected. This should come as little surprise, given that much scientific software is written by scientists with little or no training or experience in software development [6].

Many of the reported problems caused by software faults come from the physical and engineering sciences rather than the life sciences. This could be because these disciplines have a longer history of depending on computational results, a stronger culture of validation and error reporting, or simply a different approach to computational studies. However, computational work is now becoming very important in the life sciences as well – with examples of use ranging from commercial and large-scale community projects in systems biology, to the single-person projects that are so common in neuroscience. In this paper, therefore, we will focus on neuroscience software, and on the way the culture and practice of software development affects the validity and trustworthiness of the results it generates.

The most important results of our study are not our assessments of individual software projects, but the assessment criteria we have developed. We believe that these criteria can be used to understand why some projects yield more useful tools than others, and also to guide expectations about the results of software development activities. This kind of analysis may, in turn, help funders and researchers to decide how best to get the software they need, and make it easier for developers to decide which projects to work on. Our criteria could also help explain decisions about research funding, and the ease or difficulty of publishing a particular paper - phenomena that often puzzle researchers. Even if reviewers never use exactly the criteria we are proposing, we suggest that the underlying issues contribute to their decisions.

## Methods

The best starting point for a study of current practice in software development in computational neuroscience is a representative sample of software tools. We therefore reviewed two lists: one covering modeling tools from the INCF Software Center, the other a list of simulation tools curated by Jim Perlewitz [7]. To be listed, tools must have been proposed by their developers or have achieved enough visibility to be added by third parties. This means that the two lists on their own present a representative sample of the software currently available to computational neuroscientists. To complete the picture, we also examined a sample of projects from open software repositories, including SourceForge, GitHub and BitBucket.

For our review, we used information about each tool's update history, available versions, current status, together with any record of publications using the tool. The majority of this information came from the tool's primary website and source-code repository. The review was somewhat subjective and almost certainly incomplete. However, it fulfilled our purpose of range of tools available and establishing ways of categorizing them that could be useful in future evaluations. For this reason, we have not listed all the tools reviewed, or given our observations for each individual case. Instead, we present general observations that emerged from the study, using some of the tools we examined to illustrate our points.

Although we have attempted to look at each tool objectively, it is worth stressing that our study is intended to offer new perspectives, rather than a dispassionate, empirical study of a field in which we are deeply involved. As such, it is informed by our own experience as readers and reviewers of software papers, and by anecdotal accounts of problems with scientific software projects over many years.

## Results

We considered about 50 software projects, including sub-cellular simulators, simulators for large networks, and programs for interactive investigation of complex dynamical systems. All the tools were developed by or for neuroscience researchers, and were publicly available. About half of them appeared to be still maintained. Of the rest, half had clearly been abandoned. The reminder appeared to be inactive or dormant.

The criteria we found most useful can be succinctly summarized under four headings: "What", "Why", "Who", and "How".

### "What?": types of scientific software

The "What" axis concerns the end product of the development effort. How should we classify a particular software tool? What should we expect from it? Is it like a commercial product, or just some sample source code that may prove useful? Both kinds of software can benefit the community in their own way. However, researchers who expect one kind of software are disappointed when they find the other.

We suggest that scientific software can be split into four broad categories. First, there are exercises: the software produced by developers as a way of acquiring skills and testing new ideas. Second, there are the reference implementations they use to back up new models or algorithms they are publishing - software that has very different requirements from tools intended for regular use by third parties. Third, there are private tools written to address a particular problem and normally used only by a single individual or group. Finally there are public tools: fully fledged software products intended for public use. In what follows, we will consider each of these four categories in turn.

## 1. Exercises and proof of concept software

Much research software is written to test an algorithm or to advance a researcher's understanding of the ideas or data involved. In these conditions, there is little need to impose a clear separation between the model and the implementation or to write user documentation.  The most important characteristic of this category of work is that the final outcome is not the software itself. Where the software is created as a training exercise, the outcome is the trained individual. Where the software is created to explore or develop an algorithm, the algorithm should stand without reference to any specific implementation.

## 2. Reference implementations

Good examples of reference implementations can be found in the supplementary data to the classic Izhikevich neuron model [8] or the original publication of the Mersenne-Twister algorithm for random number generation [9]. One of the main characteristics of a reference implementation is that the source code should favor readability by other developers over computational efficiency. As such it does not need the kind of logging, error handling, or user documentation called for in production software. Indeed, such features may obscure the core algorithm. A concise and minimal implementation is easier to read and easier to incorporate in other tools, which provide their own logging and error checking.

## 3. Private tools

Many research groups develop and maintain private tools for simulation and for data analysis. Such tools are frequently directed at very specific problems and changes to the software may be required for each new problem. Knowledge about what the software does and how it is used may be largely unwritten and passed directly between users. The benefit of such private tools is that the development effort required is typically much lower than for general-purpose solutions.  When the research performed with such a tool is published, the software is typically described in the methods section, rather than being the focus of the publication. Descriptions are generally brief and do not incorporate much information about the testing and validation of the software.

## 4. Public tools

This category is reserved for tools that have most of the characteristics of commercial software products including broad scope, robustness, demonstrable correctness and adequate documentation. With this class of software, it should be possible for new users to undertake effective work without recourse to the original developers and without requiring modifications to the source code to address new problems within the software's intended domain of application. This requires a good internal design with a clear separation between the specification and implementation of a model, and may require scripting or plugin support for extensions. The creation of new public tools generally involves publication of papers about the software itself, including the methods involved and steps taken to validate it. Many widely used tools such as Neuron [10], GENESIS [11], and NEST [12], fall into this category.

**Unfinished, abandoned and unused tools**

All software development carries the risk that it will fail to produce anything useful. This can happen for a variety of reasons including insurmountable technical difficulties, lack of experience among the developers, bad choices early on in the project, or simply because the software, as originally conceived, is of no use to researchers. Indeed, this last case is alarmingly common, particularly among capability-driven projects, as described in the "How" section below. Software that is developed by individual researchers or students for their own purposes and never used again, can be most charitably regarded as an exercise or a proof of concept: almost certainly, the developer has gained some insight or understanding from writing it. The same cannot be said where work is delegated to in-house developers or students working on behalf of someone else. This kind of software is a waste of valuable resources. When it fails to deliver adequate results, an analysis of what has gone wrong can yield valuable lessons for the future. We will return to this issue later in this paper

## "Why?": the origins of neuroscience software

The "Why?" axis is characterized by a strongly bimodal distribution. One group of software consists of "demand-driven" projects. These arise where software is needed to solve a particular problem, and the focus is on developing a tool that can help with the research in hand as soon as possible. At the opposite end of the spectrum are "capability driven" projects. These are projects driven by technological opportunities: the kind of projects that emerge when it first becomes possible to perform a new kind of computation, or because the current fashion in software development shifts in favor of one design approach over another.

**Demand-driven software**

The characteristic of demand-driven software is that there is a user, or "customer", for the software from the very beginning. The customer could be the same person as the developer, where researchers or students write software for their own needs, or an independent party, as when research groups hire in-house developers or contract independent developers.

A good example is the neuroConstruct project, which is being developed by the Silver laboratory at UCL [13]. In this case, there is only one developer, and the group is its own customer. Several people in the group are engaged in research that is dependent upon the software being developed.

Commercial software efforts are demand-driven. Some speculative efforts may explore new technologies in the hopes of creating a market. However, investors in such projects are understandably keen to see market demand emerge as soon as possible, so they can cut their losses if it does not.

## Capability-driven software

By capability-driven software, we mean software that is developed because it might be useful, and there is funding for it. Since capability-driven software has no acute demand, it typically has no users.

Capability-driven software can occur on any scale. At an individual level, students may rewrite a perfectly adequate tool in a different language, simply because they prefer it. This can be good training, but is unlikely to yield an improved software product. On a larger scale, informatics groups are sometimes successful in attracting funding to build what they believe neuroscience researchers need. This model has been tried extensively in various countries, attracting substantial investment in the UK's e-Science program, but has often led to disappointing results. Typical problems include building the wrong thing (lack of market research before funding), a fixed-term development cycle with staff typically hired for three years, and no sustainable continuation plan. Even when the project yields useful software, the initial grant funding usually stops at just about the time when the project can be expected to attract external users.

A typical example is the NeoSim framework [14], developed under the e-Science project in connection with the US Human Brain Project. At its peak, it had 5 programmers but never attracted any users. The developers departed at the end of the grant with very little to show for their efforts, and none of the software was ever re-used. This was possible because NeoSim was a purely technology-driven proposal without a specific scientific application. Although it had the potential to grow into a useful product, the lack of demonstrable demand within the funded period made it a poor candidate for continuation. NeoSim addressed problems of connecting simulators, addressed more recently by MUSIC [15] and the Blue Brain Project [16]. In such projects, developers need almost superhuman prescience, if they are to build something useful. This is a critical issue for a number of recent large-scale projects promising new software for neuroscience research. It is closely related to the burgeoning population of empty databases and unused web applications that have been built because developers thought others would use them [17,18].

Capability-driven software is sometimes sold to funders on the grounds that it will not only help neuroscientists, but will also generate new computer science results. This proposition is based on the optimistic assumption that getting computer scientists to write brain modeling software will help them understand the brain. We know of no cases where this has actually happened.

## "Who?": software developers in computational neuroscience

The need for new software can be met in many ways. In some cases, researchers write it themselves. More often, research students work on software projects either as part of their research work, or as a means to develop the models and simulations needed for their research. In either case, the work is done by individuals whose primary motivation is the research outcome itself. For larger projects, a research group may be able to hire in-house developers who focus exclusively on the software, or they may outsource development to commercial entities. Each model has its own advantages and disadvantages. In general, there is a strong correlation between who does the work and the type of software that is produced.

## Researchers

For researchers addressing new problems in computational neuroscience, the most direct way to develop the required software is to write it themselves. Many computational neuroscientists have extensive software development experience and can write very good software. By developing on their own, they eliminate the need to communicate their ideas to a third party, and can achieve results fast. However, software development is very rarely recognized as a primary output of research positions, and often competes for time with teaching, grant-writing, supervising students, administration, and, of course, the research itself. Interestingly, researchers' ability to engage in software development seems to be inversely correlated with their requirement to engage in teaching and university administration. The NEST simulator [13] and PyNN language [19] are good examples of projects in which a significant part of the development effort comes from researchers in permanent positions. These projects also have considerable input from students, but the active involvement of full-time researchers gives them a degree of continuity and coherence that is hard to achieve by other means. Interestingly, all the lead researchers are based in continental Europe, even if some of them are from the UK and North America.

## Students

For researchers who do not have the skills, time or inclination to write new software themselves, having it written by research students is a natural (and cheap) alternative. Many research projects require new tools and, in the absence of additional funding, there is little alternative to having the software written by students. This is a challenging way of developing new software, for several reasons. First, students typically do not have professional software development experience before starting. Therefore, they must learn to write good software, as well as learning to do research. However, software skills are best learned by working among more experienced developers, and very few labs can provide this environment. This often leads to the problems described by Baxter et al. [1], with students having to make software design and architectural decisions on their own before they are really ready. Secondly, research projects are often too short for software to be written and interesting research results to be achieved with it. The outcome of standalone student projects is therefore very variable. Some students write great software, but by their own admission, others spend their time learning by trial-and-error. In the latter case, the best approach is to extract any good ideas and start again from scratch. Finally, writing software for other users reduces the time available for research, and is very risky for researchers wishing to pursue a research career. This issue is particularly problematic where researchers hire students for projects with a very large software development component, because they need the software for their own research. The students have very little chance of developing research careers, and are in effect serving as fixed-term contractors, paid a fraction of the market rate.

**In-house developers**

An alternative to using research students for software development is to hire software engineers, whose primary goal is the development of good software to be used by other people. This has the advantage that the developers' actual work is well aligned with their own objectives and job descriptions. Because they work only on the software, they are also able to complete many of the associated tasks such as developing documentation, examples and tutorials, which are essential for practical software products, but which are not on the critical path to generating research results. Some of the most successful long-term software systems in use in computational neuroscience, including both Neuron and Genesis, have benefited from in-house developers. However, it should be mentioned that, in both cases, the developers had also been involved in original research. At present, the greatest challenge for software development by in-house developers is funding. With almost no permanent university positions for this kind of work, developers depend on successive short-term contracts and rely on their principal investigators (PIs) for continued funding. For PIs, keeping a good in-house developer can be very difficult, as any break in funding will force them to find work elsewhere, and it may be hard to hire them back afterwards.

**Outsourced developers**

A potential solution to the problem of providing continuity for in-house developers is to outsource the work to commercial or non-profit organizations that undertake software development for a number of clients. In principle, such organizations can even out the flow of funds from different projects, and provide their developers with a more secure career path than in-house developers. Perhaps the biggest difficulty with this model is finding software engineers with sufficient qualifications in specific domains of research.

## "How?": Development models for neuroscience software

The means by which scientific software is developed vary according to the needs of the project, funding sources and the interests of those involved. Observations of current projects suggest three broad categories. First is the Heroic Model, where one developer works on a piece of software over several years. Second is the Collaborative Model, in which researchers from different groups pool their resources to develop and maintain a piece of software. Finally, in the Outsourced Project Model, a research group contracts an independent software developer to write a particular piece of software.

## The Heroic Model

This was the most common development model in the projects we reviewed. A researcher begins writing software to address a particular problem. Over time, the software accumulates features and the researcher decides to share it with others. In some cases, this point is reached at the end of PhD theses, when the researchers think that they will not be able to continue to develop the software themselves and release it to the community. In other cases, a researcher continues development, and other members of his/her research group get involved in using or extending the software, which remains entirely within the group.

The biggest challenge with the Heroic Model is that such tools rarely reach the maturity and completeness required to constitute a public tool. Notable exceptions are Brian [20] and Topographica [21], which have gained momentum, and partially transitioned into other development models. However, many of the Heroic tools we examined have a single developer and no apparent ongoing activity. The other main weakness of the model is the emphasis on a single developer. This creates a single point of failure, with no means to ensure continued use or development of the software if the original developer is no longer available. In brief, the Heroic Model lacks adequate mechanisms to motivate and reward other developers for extending and supporting the original work for the benefit of the community.

It is natural for research software projects to begin with the Heroic Model. One of the main challenges with neuroscience software is to promote the transition of the best such projects into more sustainable models. A notable exception is the NEURON modeling package, by far the most widely cited of the tools examined in this study,. First developed by Michael Hines in the late 1970s, NEURON is still developed and maintained by its original author, who has received uninterrupted NIH funding from 1978 to the present. This shows that the Heroic Model is capable of delivering long-term solutions, albeit under rather exceptional circumstances. A more typical situation may be that of Genesis 3, where the lead developer was forced to seek work in the private sector due to lack of continuity in funding.

## The Collaborative Project Model

Collaborative projects arise where a collection of individuals or research groups with similar requirement pool their resources, with each participant focusing on aspects of the project relevant to his or her own work. In this model, the participants benefit from a shared core code-base, shared infrastructure and the increased visibility that comes from being part of a larger effort.

One of the best examples of a collaborative project is the neural simulation tool NEST, developed by the NEST Initiative. NEST started in 1995 under the name SYNOD [22] and has been under active development ever since. At the time of writing, NEST has over 10 developers working on different parts of the software. Having users contribute what they need for their own work ensures that features are implemented as and when they are needed, and that each new feature is tested out on real scientific problems before being released to the wider community.

Although this type of development is rare in neuroscience, it is much more prevalent in related disciplines. In particular, as De Schutter noted in this journal [23], systems biology is currently in a very different situation from computational neuroscience. Demands arising from the flood of data from increasingly industrialized processes in systems biology have lead to large-scale collaborative software projects. The active development of new software for which there was a clear community need has enabled the development of and support for community standards such as MIASE and SBML. De Schutter contrasts this with the situation in computational neuroscience, where much "computational neuroscience software is shackled by legacy code" [23,24]. The focus on large-scale projects in systems biology makes it possible to employ specialists for different roles within a project. In particular, systems biology can employ scientific programmers who are not expected to double up as researchers, and for whom there is a credible, long-term career path in providing the software engineering component of a much larger activity.

## The Outsourced or Market Model

In this model, researchers in need of software contract an independent company or individual to write it. The development of PSICS [25] was outsourced at a fixed price by a research group that needed the software for its research. In this case, documentation and validation amounted to 40% of the total cost, with the core functionality carefully defined to produce a tool that the researchers could use on their own. After the initial work, two research groups contracted the original developers for additional work to meet their specific requirements. All the outputs are fully documented and open source.

This model has yet to be used extensively in neuroscience, but the example of PSICS suggests that this could be at least partly due to the lack of suitable organizations to outsource to, rather than a lack of interest from the community. In principle, this model offers advantages to both sides. Researchers can negotiate a fixed-price contract to be paid on delivery of working, validated software. Small projects can be accommodated and the original developers are more likely to be available to carry on, when more work is required and additional funding is available. For software developers with an interest in science, such organizations could offer a stable career path in a single location, while working on a succession of different projects and with the kind of close contact with other software engineers that is essential for effective professional development.

## The Community Engagement Model

This model has a long history in other areas of software development, for example, the Linux operating system has a large community of independent developers. However, it is a relatively new approach for neuroscience. One of the best recent examples is the OpenWorm project (openworm.org) where a community of developers got together to create a biomechanical and physiological simulation of *C. elegans*. Most of them have no specific scientific training, but they are able to read the literature and implement the models. Interestingly, the project's success in developing working software is making it increasingly attractive to researchers, who contribute their experimental data and offer their expertise in computational modeling. Although it is rather early to assess scientific outcomes, OpenWorm has already jumped some of the hurdles facing projects that originate in the scientific community. It has a large and active community of developers. The software itself is of a high standard, and there is a healthy balance between core development and the development of specific products, including visualization tools, documentation and an accessible web presence.

# Discussion

After inspecting a wide range of software projects, we find that these projects can be usefully categorized by the answers to four key questions: "What", "Why", "Who", and "How" . We suggest that this scheme usefully captures the main factors determining the long-term success of a software project, and consequently the value it represents to the research community, and the advancement of neuroscience research.

However, there is one additional assessment criteria that we have not been able to consider so far: we have not been able to assess whether individual pieces of software are correct, in the sense that they correctly implement the models they are intended to implement. The reason for this omission is that in most cases, the necessary information does not exist. In the past, when computational modeling was something of a fringe activity, this situation could be tolerated. Today, however, simulations and other computational results are playing an increasingly important role in science, political decision-making, and society as a whole [26, 27], and researchers are becoming increasingly aware of the critical influence of software quality on the sustainability on their research. In these conditions, scientific software can no longer be the private affair of the scientists who develop it.

## Assessment of scientific software

Given the growing importance of scientific software, the community needs ways of assessing whether a particular exercise, reference implementation, or tool is fit for its intended purpose. The first threshold a software tool must cross is that it must be "research-ready": ready to be used for research by the person who wrote it. This means that the developer can be confident that the results it produces are correct. For simple scripts, this may be achieved by inspection, but in most cases it will be necessary to apply the tool to test models for which there are known analytic solutions, or to models generated by other tools. With the exception of the three Rallpack tests [28] for single cell simulators, computational neuroscience has very few standard tests. In comparison, the Systems Biology Markup Language (SBML) test suite comprises more than a thousand test models, complete with detailed descriptions and expected results for comparison. A number of projects are under way to address these issues (see for example. http://opensourcebrain.org). The late development of comprehensive test suites in computational neuroscience can be largely attributed to the absence of shared model description formats [29]. With the emergence of NeuroML [30] for single cell models and PyNN for networks, the coming years will hopefully see major improvements.

Software that is research-ready may still not be suitable for use by other researchers. As well as being correct, wider usage requires that it is accessible and usable. For example, it should have comprehensive documentation and examples, as well as sufficient error handling and reporting functionality to enable users to trace problems with their models without recourse to the source code. Below, we propose checklists for assessing whether a particular tool meets these criteria.

## Creation of public tools

One of the stated aims of many projects involving software development is to produce software that will be of use to other researchers. In our terminology, this entails creating a public tool. Of the criteria considered here, the "Why?" axis has the clearest correlation with the eventual emergence of a public tool from a software development activity. Somewhat obviously, unless the implementation of a demand-driven project is so bad that the software cannot be used at all, these projects almost always find at least some use in research. Conversely, capability-driven projects only find a use if what has been developed happens to coincide with a research community need. Much highly specialized software never has this good fortune.

The "Who" axis is also important. Should software be written by scientists, or delegated entirely to professional software engineers? On the one hand, core algorithms cannot be developed without the involvement of scientists. This means scientific software inevitably has a close link to the latest research. However, Wilson [6] found that very few researchers are familiar with best practice in software development. Our own observations suggest that this situation has changed somewhat over the last seven years. With the explosion of open source activity on GitHub and BitBucket, and the increasing use of community sites such as StackOverflow to discuss design and development practices, it is now much easier for developers to keep up with new practices, even when they are not working in a software company. Indeed, the preponderance of short-term projects may act in scientists' favor, since they are able to adopt new tools as they emerge, rather than being tied to long-term corporate structures.

However, the process of turning a private tool into a public tool is very demanding in terms of programming, testing, and documentation. According to some estimates, in fact, this step requires up to nine times the effort needed to develop a private tool [31]. Individual developers and laboratories do not have the resources or skills to transform a private tool into a public one, and to handle its subsequent distribution and user support. This task could, however, be handled by spin-off companies or other commercial entities. This system works well in experimental biology, where many of the companies now supplying laboratory equipment have their origins in research laboratories.

Another problem for the users and developers of scientific software is that funding systems, and the career paths of research students and junior researchers tend to favor the development of new tools over the extension and maintenance of existing ones. This explains why a high proportion of early stage projects in our sample are no longer supported. Taken together, these observations highlight the need for funding and projects that fill the gap between innovative single-developer projects and research-ready software applications. In neuroscience there is currently an ample supply of early stage projects, but almost no mechanism for turning them into useful public tools.

## Sharing of software

Sharing software is widely considered an important step in improving quality in the computational sciences [32,33]. While this is certainly true, it is also important to realize that there are different reasons why software should be shared. Accordingly, there has to be a range of different standards for shared software.

Software used in research studies should be made freely available when such studies are published, if not before. However, making software available should not be confused with asserting that the software is ready and usable by other researchers. Reference implementations of a novel algorithm or model, such as those provided by Izhikevich [8] and Matsumoto [9], may be of interest primarily for other software developers who are developing their own implementations of the models. A reference implementation may also be useful for generating test cases to compare with other tools, but not suitable for running simulations on the scale needed by a specific research problem. Simulation scripts and other iterative development work used by a single group in pursuit of a research problem may not be sufficiently general or well documented for other researchers to use it "as-is". It is nevertheless important to have them available for future examination, if their results are challenged or if other groups wish to implement exactly the same configuration as a reference point for a new study.

We see three distinct motives to share software with the scientific community:
1. To allow other researchers to evaluate and understand how a particular

numerical or simulation result was obtained. This is the case for most model and simulation code. We can say here that the code is shared for 'reading'.

2. To allow researchers to develop their own implementation of a model or algorithm, based on a published reference implementation. The reference implementations of the Izhikevich neuron [8] constitute such a case. However, they also illustrate the problem of 'error propagation'. Although the implementation provided in the original paper has known numerical problems [34], it is still often regarded as the 'correct' implementation and is used in many applications. A possible solution is to introduce curated reference implementations of models and algorithms, similar to those found in the systems biology community or in the well-known Boost library (www.boost.org). In such cases, code is shared for 'reading and writing'.

3. Finally, software is shared or, better still, published so that other researchers can use it as a tool for their research. In most cases, the code will be rather complex due to user-interfaces, error-handling and other infrastructure code, so the general user will not be able to extract and understand its core algorithms. In this scenario, the software is mostly 'used', rather than read or modified.

Each of these three "use cases" requires different quality standards for when the software is shared. The lowest level is what we call "review-ready". The source code is prepared and documented, so that reviewers and scientists can understand its main algorithms. The next highest quality standard is "research-ready", meaning that the software is sufficiently well tested and documented that its research results can be trusted. Finally, the highest quality standard is what we call "user-ready". At this level, the software is sufficiently well tested and documented that researchers who are not familiar with the source code can use it to generate research results that can still be trusted.

## Recommendations

These considerations have led us to formulate a number of suggestions for improving scientific research that is heavily dependent on software. These suggestions, which may be of interest to researchers, funders, and software developers, are presented as points for consideration, rather than as definitive recommendations. The only firm recommendation is that the problem needs to be recognized and addressed.

1. Not all software development effort can or should lead to the creation of public tools. There are differences between proofs of concept, private tools and public tools. Funders should not expect to pay for proof of concept work and have the code released as a public tool. Developers should not expect to publish a paper about a private tool as though it was a public one. Checklists such as those presented below can make it easier to decide what category a tool is in.

2. Journals should formalize their policies on what is required for different categories of publication, including papers about novel algorithms with proof-of-concept software (reference implementations), research papers where the results are generated by software, and papers about new public tools. A blanket requirement simply to make the code available risks confusing the picture, and making it hard for readers to distinguish between different sorts of software. We are not against developers being open about their work and making their source code easily accessible, indeed, we are very much in favor of this approach. Rather, our concern is that this kind of visibility can too easily be confused with a suggestion of "research-readiness". As a starting point, we suggest that:

    a) Papers about software should only be published when the software meets the criteria for a public tool.

    b) Where research papers depend on software, the software should either be an existing public tool, or reviewers should have access to the code and verify that it meets the standards of a private tool. Ideally, there should be separate peer-review for such software (see item 4).

    c) For proof-of-concept work, the ideas should be able to stand on their own and papers should make minimal mention of the specific implementation. However, it is often useful to make an implementation available to facilitate adoption, for example, as Gillespie [35], Izhikevich [8] and Matsumoto [9] did. In this case, the software submitted should meet the criteria for "review-readiness" laid out in the first checklist below.

3. When considering papers about public tools, at least one reviewer should be asked to look only at the software, perhaps using checklists such as those below. If, as at present, reviewers are asked to consider the software as well as reviewing the rest of the paper, it is almost inevitable that consideration of the software will be a secondary concern, at best. Developing a review model that includes consideration of code opens up a new pool of well-qualified reviewers (developers of other scientific software) who are rarely involved in the review process at present.

4. Rather than leaving software review solely to journal reviewers, the community could organize some form of software certification, perhaps under the aegis of the International Neuroinformatics Coordinating Facility (INCF). This could operate independently of the journal review process, would lessen the burden on reviewers, and might be able to offer a more standardized assessment of user-readiness in public tools. It would also offer a mechanism for researchers who are not software specialists, to have expert involvement in and assessment of the projects they run. A role model for this sort of software review and certification could be the peer-reviewed C++ library Boost (see www.boost.org).

5. Funding for software development should be mediated by the intended beneficiary — the scientist with research to do — rather than flow directly from funder to developer. The latter model has consistently failed to produce the tools that users actually want. Although they are not the focus of this study, similar arguments can be made for databases and other repositories that have generally remained unpopulated when their development was not driven by the end users themselves [17,18]. While this approach may slow the development of new software, it ensures that prospective users become involved in the design and development of the tool at an early stage, minimizing the risk of creating the "wrong kind of tool".
6. Publications involving novel software or algorithms should, wherever possible, include reference models and data in a standardized format, after the manner of the Rallpack tests [28]. These reference models should then be used to verify that future tools correctly implement existing models (item 3 in the second checklist below). In this way, even if new implementations start from scratch, their scope can grow incrementally, instead of just repeating the same errors as earlier projects.

## Checklists

We propose three checklists to help assess whether a particular piece of software is "review-ready" (meaning that it is suitable for examination by a reviewer in conjunction with the publication of a novel algorithm), "research-ready" (a private tool that is suitable for generating publishable results), or "user-ready" (constituting a public tool that is both "research-ready" and adequate for use by independent third parties). In each case, all the related statements should be true for the software to qualify.

In compiling the checklists, our intention has been to establish a minimal pragmatic set of requirements that can be realistically achieved, and which can help to alleviate some of the basic problems that plague scientific software today. The checklists do not address issues of software quality in terms of systems architecture or coding. Although of obvious importance, such considerations are beyond the scope of this paper. They are, however, covered by software life-cycle models, such as Tribits, developed by the Trilinos Project of the Sandina National Laboratories [36].

### Criteria for proof-of-concept software to be "review-ready"

1. Software written in a compiled language is easy to compile and runs without crashing.
2. Software written in an interpreted language is easy to install and runs without warnings and error messages.
3. The software favors directness and simplicity over computational efficiency, where the former provides a clearer demonstration of the

algorithm.
 4. It contains enough commentary to easily relate sections of the code to the written presentation of the algorithm.
 5. It comes with simple test cases that can be run easily.

**Criteria for private tools to be "research-ready"**

 1. Software written in a compiled language is easy to compile and runs without crashing.
 2. Software written in an interpreted language is easy to install and runs without warnings and error messages.
 3. The software offers basic error handling and diagnosis.
 4. Previous versions are archived and readily available, so that results produced with a previous version can be regenerated.
 5. The software comes with test cases where analytic or previously computed results are known.
 6. The software implements any relevant consistency checks, such as conservation of mass or charge.

Additionally, publications containing results generated with private tools should include test cases for which the tool's behavior is already known, or can be independently predicted, demonstrating in this way that the model that has been implemented is indeed the model that was intended.

**Criteria for public tools to be "user-ready"**

 1. The software meets all the criteria for a private tool.
 2. The software comes with implementations of previously published models demonstrating that, the software generates correct results, at least for the cases provided.
 3. There is comprehensive user- and developer documentation that enables qualified individuals to work with the software without recourse to the developers.
 4. The user interface to the tool, whether graphical or command-based, conforms to usability and design norms for software of this type.

## Conclusion

We have examined a wide range of software that has been created for use in neuroscience research. We find that some of the dismay of other authors at the current state of affairs can be attributed to a misunderstanding as to why particular software was created and what can reasonably be expected of it.

Writing bad software is an inevitable step in the professional development of anyone who will eventually write good software. Much of this work involves solving problems that have already been solved. To dismiss this as reinventing the wheel is like arguing that pianists shouldn't learn to play pieces that other musicians can already perform better. The difference, of course, is that trainee musicians do not publish recordings of their work. This, we suggest, is where the real problem lies. Much scientific software is, in effect, written by early stage trainee software engineers. Unlike the cacophony made by music students, much neophyte software output gets recorded for posterity as though it was publishable work. This leads to the unusable programs and labyrinthine codebases that so distressed Baxter et al. [1]. From this perspective, the problem is not that trainee software developers write bad software, but that this software is misrepresented to others as being ready for use in solving scientific problems, or as a basis for extension by other developers.

Based on this observation, we suggest that the best way to improve the situation is to recognize the different types of software development activity within science, and adjust expectations accordingly. We hope that by formalizing the progression of software, from exercise or proof-of-concept code, to useful multi-user tools, including notions about what is publishable, and when software becomes ready for use in a scientific investigation, much of the current confusion around software quality, validity and suitability for publication may be avoided. Inevitably, such a realignment of expectations will result in much existing software being reclassified as not yet "research-ready". However, such an outcome can only be beneficial in driving the creation of truly reliable research tools, and improving the credibility of software-dependent research results.

## Acknowledgments

We thank Hans Ekkehard Plesser and Markus Diesmann for valuable comments on earlier versions of the manuscript. Thanks also to Richard Walker and Guy Willis for proofreading the final version of the manuscript.